\documentclass[a4paper,11pt]{article}
\usepackage{pos}
\usepackage{diagbox}
\usepackage{bbm}
\usepackage{amsmath}
\usepackage{graphicx}
\usepackage{subcaption}
\usepackage{float}
\usepackage{simpler-wick}
\PassOptionsToPackage{hyphens}{url}
\usepackage{hyperref}

\title{{\normalsize {\rm \hfill{HU-EP-24/17-RTG} }}\\
\vspace{0.8cm}
An update on the supersphere non-linear sigma model on the lattice}
\author*{Ilaria Costa}
\author{Valentina Forini, Agostino Patella, Johannes Heinrich Weber}

\affiliation{Institut f\"ur Physik, Humboldt-Universit\"at zu Berlin, IRIS Adlershof, Berlin}

\emailAdd{ilaria.costa@physik.hu-berlin.de}
\emailAdd{valentina.forini@physik.hu-berlin.de}
\emailAdd{agostino.patella@physik.hu-berlin.de}

\emailAdd{johannes.weber@physik.hu-berlin.de}
\FullConference{%
	European network for Particle physics, Lattice field theory and Extreme computing,\\
	11-15 September, 2023 \\
	Berlin, Germany 
}

\abstract{	
	We consider the discretized version of the sigma-model with supersphere target space $OSp(N+2m|2m)/OSp(N+2m-1|2m)$ introduced in \cite{Costa:20232Z} and present a preliminary numerical study of bosonic and fermionic two-point functions for the cases $OSp(3|2)$ and $OSp(5|2)$. We observe consistency with the expectations of this supersymmetric setup and discuss the sign problem.
}

\begin{document}
	\maketitle

	\section{Introduction}
	Two-dimensional $O(N)$ non-linear sigma models are renormalizable \cite{Brezin1976}, exactly solvable theories \cite{Zamolodchikov1979} and have many applications, from statistical mechanics to their use as QCD toy models \cite{Novikov:1984ac,DAdda:1978dle,Polyakov:1975rr,Pelissetto:2000ek}. Consequently, they have been an object of thorough study via lattice QFT methods. 
	
	A simple supersymmetric extension of the $O(N)$ non-linear sigma model is the sigma model on the supersphere $OSp(N+2m|2m)/OSp(N+2m-1|2m)\equiv S^{N+2m-1|2m}$, $N,m$ non-negative integers.
	A number of analytic properties of it -- such as the spectrum of local operators at the renormalization group fixed-points, their integrability properties and their integrable deformations -- have been studied in~\cite{Read:2001pz, Saleur:2001cw, Saleur:2003zm, Babichenko:2006uc,Mitev:2008yt, Cagnazzo:2014yha, Alfimov:2020jpy}. 
	This setup  provides a simple ground to gain experience in analyzing lattice quantum field theories of two-dimensional sigma models on supersymmetric target spaces. These  have important applications in various areas, ranging  from statistical mechanics~\cite{Parisi:1980in, Gruzberg:1999dk, Quella:2013oda, Zirnbauer:2018ooz} to, notably, string theory and the AdS/CFT correspondence~\cite{Maldacena:1997re,Metsaev:1998it}. A lattice discretization of string worldsheet models in AdS, however, may present non-trivial challenges~\cite{Forini:2016sot,Forini:2016gie,Bianchi:2016cyv, Bianchi:2019ygz, Bliard:2022kne,Bliard:2022oof}.

In this paper, we work on the discretized version of the model introduced in \cite{Costa:20232Z}  and present preliminary  numerical results for the specific cases $OSp(3|2)$ and $OSp(5|2)$.

This paper is organized as follows. In Section 2, we provide a brief overview of the model and its key properties. In Section 3, we discuss the relation between $n$-point functions in the $OSp(N+2m|2m)$ sigma model and those evaluated in any other such model with different, positive integer $m$ ($m'$ below). 
In Section 4 we describe the simulation setting,  in Section 5 we outline our numerical approach for computing the two-point functions and discuss the features of the emerging sign problem. Conclusions are drawn in Section 6. 

	\section{The model}
	The $OSp(N+2m|2m)$ non-linear sigma model is defined in terms of a superfield $\Phi$ that maps a two-dimensional flat space to the supersphere $S^{N+2m-1|2m}$, which is the target space of the model.
	The superfield $\Phi$ can be decomposed in its bosonic and fermionic components
	\begin{gather}
		\Phi\equiv(\phi^1,\ldots,\phi^{N+2m},\psi^1,\ldots,\psi^{2m})\ .
	\end{gather}
	The first $N+2m$ components are bosonic and the remaining $2m$ are fermionic. We introduce the following bilinear form:
	\begin{gather}
		\Phi\cdot\Phi^\prime=\phi^T\phi^\prime+\psi^T\,J\,\psi^\prime\ ,
	\end{gather}
	where $J$ is the $2m\times2m$-dimensional canonical symplectic matrix
	\begin{gather}
		J=\left(\begin{array}{cc}
			0& \mathbbm{1}\\-\mathbbm{1}&0
		\end{array}\right)\ .
	\end{gather}
	For the superfield $\Phi$ to live on the supersphere, it has to satisfy the constraint:
	\begin{gather}
		\Phi\cdot\Phi=\phi^T\phi+\psi^T\, J \,\psi=1,
	\end{gather}
	The path integral is defined in the following way:
	\begin{gather}
		Z_{(N+2m|2m)}
		=
		\int D^{(N+2m-1|2m)}\Phi \ e^{-S_{(N+2m|2m)}(\Phi)}
		\label{eq:def:Z}
	\end{gather}
	where action and measure are
	\begin{gather}
		\begin{split}
			S_{(N+2m|2m)}(\Phi)&=\frac{1}{g}\sum_{x,\mu}a^2\partial^f_{\mu} \Phi_x \cdot \partial^f_{\mu} \Phi_x
			=\frac{2}{g}\sum_{x,\mu}a^2\left[1-\phi^T_{x+\mu}\phi_x-\psi^T_{x+\mu}\,J\,\psi_x\right],\\
			D^{(N+2m-1|2m)}\Phi &=\prod_{x}\delta\left(1-\phi_x^T\phi_x-\psi^T_xJ\psi_x\right) d\phi_x d\psi_x,
		\end{split}
		\label{inv_meas}
	\end{gather}
	the sum over $\mu$ is over the two directions on the worldsheet and $\partial^f_\mu$ is the discrete forward derivative in the direction $\mu$.

	Both lattice discretized action and  path integral are invariant under the action of the supergroup $OSp(N+2m|2m)$, whose algebra can be represented by the super-matrix
	\begin{equation}
		S=	\left(\begin{array}{cc}
			S_{\phi \phi} & S^T_{\psi \phi}\, J\\
			-S_{\psi \phi} & S_{\psi \psi}
		\end{array}\right), 
	\end{equation}
	where $S_{\phi\phi}$ is an element of the $\mathfrak{so}(N+2m)$ algebra, $S_{\psi\psi}\in \mathfrak{sp}(2m,\mathbb{R})$, while $S_{\psi\phi}$ is an anticommuting  $2m\times N$-dimensional matrix. The field coordinates transform as
	$\delta\Phi=S\,\Phi$, in other words $\delta \phi=S_{\phi\phi}\phi+S^T_{\psi\phi}\,J\,\psi$, $\delta \psi=-S_{\psi\phi}\phi+S_{\psi\psi}\psi$. 
	The $OSp(N+2m|2m)$ symmetry group mixes the bosonic and fermionic degrees of freedom and thus can be interpreted as an internal supersymmetry of the target space, the supersphere. However it is not a spacetime symmetry on the flat 2d worldsheet. We point to the following properties of the partition function:
	\begin{itemize}
		
		\item For $m=0$ and $N>1$, the supersphere non-linear sigma model reduces to the $O(N)$ non-linear sigma model;
		
		\item The partition function of the $OSp(N+2m|2m)$ and $OSp(N+2m'|2m')$ non-linear sigma models are equal for any integer value of $m,m'>0$ and $N>0$;
		
		\item for $N > 0$, the partition function $Z_{(N+2m|2m)}$ is identical to the partition function of the purely bosonic $O(N)$ non-linear sigma model. This partition function is strictly positive for any real value of $g^{-1}$.
	\end{itemize}
	
	The $OSp(N+2m|2m)$ non-linear sigma model for $N+2m \ge 1$ can be proved to be renormalizable at all orders in the perturbative expansion with the lattice regulator, following the strategy of \cite{Brezin1976}. The non-linear realization of the $OSp(N+2m|2m)$ symmetry has strong implications on the form of divergences in perturbation theory, and the Ward-Takahashi
	identities constrain the form of possible counterterms, whose coefficients can be calculated as
	a function of only two renormalization constants - the coupling constant $Z_g$ and a unique field renormalization $Z_\Phi$.
	\section{Equivalence of the correlation functions}
	
	The following identity of $n$-point functions holds
	\begin{gather}
		\langle
		\phi^{a_1}_{x_1} \cdots \phi^{a_p}_{x_p}
		\psi^{\alpha_1}_{y_1} \cdots \psi^{\alpha_r}_{y_r}
		\rangle_{(N+2m|2m)}
		\nonumber \\ =
		\langle
		\phi^{a_1}_{x_1} \cdots \phi^{a_p}_{x_p}
		\psi^{\alpha_1}_{y_1} \cdots \psi^{\alpha_r}_{y_r}
		\rangle_{(N+2m'|2m')}
		\ , 
		\label{eq:equi:identity}
	\end{gather}
	provided that $a_k \le \min\{N+2m,N+2m'\}$ and $\alpha_k \le \min\{2m,2m'\}$, where $m'$ is any non-negative integer. We will briefly show how to get this result in this section.
	The $n$-point correlators are defined as
	\begin{gather}
	\langle \phi^{a_1}_{x_1} \cdots \phi^{a_p}_{x_p} \psi^{\alpha_1}_{y_1} \cdots \psi^{\alpha_r}_{y_r} \rangle = \frac{1}{Z_{(N+2m|2m)}} \frac{\delta^n Z\left[K,\eta\right]_{(N+2m|2m)}}{\delta \phi^{a_1}_{x_1} \cdots \delta \phi^{a_p}_{x_p} \delta \psi^{\alpha_1}_{y_1} \cdots \delta \psi^{\alpha_r}_{y_r}} \Bigg|_{K,\eta=0},
	\label{eq:def:n-pt}
	\end{gather}
	where $K$ and $\eta$ are the sources for the bosonic and fermionic fields, respectively. The rapid decay of the exponential in \eqref{eq:def:Z} at infinity in the bosonic variables allows the use of the following representation of the delta function
	\begin{gather}
		\delta( \Phi(x) \cdot \Phi(x) - 1 )
		=	\lim_{\epsilon \to 0^+} \frac{a^2}{2 \pi g} \int d L(x) \ e^{- i \frac{a^2}{g} \{ L(x) \Phi(x) \cdot \Phi(x) - L(x) - i \epsilon |L(x)| \} }
		\ ,
		\label{eq:equi:delta}
	\end{gather}
	Once the representation of the delta function in \eqref{eq:equi:delta} is used, the integrals over $\phi$ and $\psi$ generated from the functional derivative in \eqref{eq:def:n-pt} become Gaussian and can be explicitly calculated. The only non-vanishing Wick contractions are
	\begin{gather}
		\wick{\c\phi_a(x) \c\phi_b(y)}
		=
		\frac{ 2g \delta_{ab} }{ a^2 ( - \hat{\Box} + iL ) }(x,y)
		\ , \\
		\wick{\c\psi_\alpha(x) \c{\psi}_\beta(y)}
		=
		\frac{ 2g J_{\beta\alpha} }{ a^2 ( - \hat{\Box} + iL ) }(x,y)
		\ .
	\end{gather}
	In particular, the expectation values in \eqref{eq:equi:identity} do not vanish only if $p$ and $r$ are even. Assuming both are true, we get
	\begin{gather}
		\langle
		\phi_{a_1}(x_1) \cdots \phi_{a_p}(x_p)
		\psi_{\alpha_1}(y_1) \psi_{\beta_1}(z_1) \cdots \psi_{\alpha_q}(y_q) \psi_{\beta_p}(z_p)
		\rangle_{(N+2m|2m)}
		\nonumber \\ \qquad =
		\sum_{\sigma \in \Sigma_p} \sum_{\tau \in \Sigma_q}
		C_{(P|2Q)}(x_{\sigma(1)},\dots,x_{\sigma(p)},y_1,z_{\tau(1)},\dots,y_q,z_{\tau(q)})
		\nonumber \\ \hspace{2cm} \times
		\left[ \frac{1}{2^{p/2} (p/2)!} \prod_{i=1,3,\dots,p-1} \delta_{a_{\sigma(i)},a_{\sigma(i+1)}} \right]
		\left[ \text{sgn}(\tau)\prod_{i=1,2,\dots,q} J_{\alpha_i,\beta_{\tau(i)}} \right]
		\ ,
		\label{eq:equi:n-pt}
	\end{gather}
	where $\Sigma_n$ is the set of permutations of the first $n$ positive numbers, $\text{sgn}(\tau)$ is $+1$ (resp. $-1$) if the permutation $\tau$ is even (resp. odd), and the functions $C_{(N+2m|2m)}$ are defined by
	\begin{gather}
	C_{(N+2m|2m)}(x_1,\dots,x_n)
	=
	\frac{1}{Z_{(N+2m|2m)}} 
	\left( \frac{a^2}{\pi g} \right)^{\frac{-N+2}{2}V}\\ \nonumber\times
	\lim_{\epsilon \to 0^+} \int DL \ 
	e^{\frac{1}{g} \sum_x a^2 \{ i L(x) - \epsilon |L(x)| \} } 
	\det \{ - \hat{\Box}+ iL \}^{-\frac{N}{2}}\prod_{i=1,3,\dots,n-1} \frac{ 2g}{ a^2 ( - \hat{\Box}+ iL ) }(x_{i},x_{i+1})
	\ .
	\label{eq:equi:C}
	\end{gather}
	Since the partition function depends only on $N$, the $n$-point functions also are independent of the value of $m$ or $m'$.
	Notice that this function does not distinguish between bosonic and fermionic components. This is a consequence of the fact that the only differences in the fermion and boson Wick contractions are in the Kronecker delta and $J$ and possible minus signs due to the anticommutation of fermions.

\section{Simulation setting}
	We will now briefly review the discretized setting of the model used for our simulations. All the details of the calculations can be found in \cite{Costa:20232Z}. We restrict to the case $m=1$, i.e. with only two fermionic degrees of freedom, and consider periodic boundary conditions.  

	To run simulations of the theory, we have to manipulate the partition function to integrate out the fermion fields. First, the delta function in \eqref{inv_meas} can be integrated out if we impose $\phi$ to be expressed in the following way:
	\begin{gather}
		\phi^a_x=(1-\psi^1_x\psi^2_x)\varphi^a_x.
	\end{gather}
	The new variables $\varphi$ satisfy the constraint $\varphi^T\varphi=1$. This rescaling will give rise to interaction terms that are quartic in the fermionic fields. With the introduction of auxiliary fields $A_{\mu}$ via a Hubbard-Stratonovich transformation, the action can then be made quadratic in the fermionic fields.\\
	We finally consider the following discretized effective action for the theory:
	\begin{gather}
		\mathcal{S}_{\mathrm{eff}}=\sum_{x}\left[\sum_{a,\mu}\frac{2}{g}\left(-\varphi^a_{x+\mu}\varphi^a_{x}+\frac{1}{2}A_{x\,,\mu}^{a\,2}\right)+\sum_y\psi^1_x\mathcal{K}_{x,y}\psi^2_y \right].
		\label{final_eff_action}
	\end{gather}	
	The symmetric matrix $\mathcal{K}$ is defined as
	
	\begin{gather}
	\begin{split}
	\mathcal{K}_{x,y} = & N\,\delta_{xy} + \frac{2}{g}\sum_{a,\mu}\left[\varphi^a_{x}\left(\varphi^a_{x+\mu}+\varphi^a_{x-\mu}\right)\delta_{xy}+(A_{x\,,\mu}^{a}+A_{x-\mu\,,\mu}^{a})\varphi^a_{x}\delta_{xy}\right]\\&-\sum_\mu\frac{2}{g}(\delta_{x-\mu,y}+\delta_{x+\mu,y}).
	\end{split}
	\label{eq:K_def}
	\end{gather}
	The first term comes from integrating out the delta function in \eqref{inv_meas} and thus is independent of $g$.
	We can integrate the fermionic fields, leading to the following partition function: 
	\begin{gather}
		\mathcal{Z}=\int\prod_{x}dA_xd\varphi_x\,\delta(\varphi^T\varphi-1)\,e^{-\mathcal{S}_{\text{bos}}}\;\det\mathcal{K}.
		\label{final_path_int}
	\end{gather}
	Since $\mathcal{K}$ is a real matrix, its determinant is real. However, it is not generally positive and we will see the emergence of a sign problem in the simulations, an issue that will be analyzed in more detail in the next section. The final effective action that we have used for numerical simulations is then obtained by re-exponentiating the modulus of the determinant.
	\begin{gather}
		\mathcal S_{\text{simul}}=\sum_{x,a,\mu}\,\frac{2}{g}\left(-\varphi^a_{x+\mu}\varphi^a_{x}+\frac{1}{2}A_{x\,,\mu}^{a\,2}\right)+\sum_{x,y}\,\chi_x^T(\mathcal{K}^2)^{-1}_{xy}\,\chi_y,
		\label{action_psuedo_f}
	\end{gather}
	where $\chi$ is a real pseudofermion. The sign of the determinant of $\mathcal{K}$ is then taken into account in a reweighting factor. Once $R$ is generated, we compute all the observables using the reweighting factor. 
	
	For the simulations, we have worked with a standard Hybrid Monte-Carlo~\cite{Duane:1987de}. We have chosen the Molecular Dynamics Hamiltonian 
	\begin{gather}
		\mathcal{H}=-\sum_{x}\left[\frac{1}{2}(\pi^a_x)^2+\frac{1}{2}\sum_{\mu}(p^{a}_{x,\mu})^2\right]+\mathcal{S}_{\text{simul}}(\varphi,A),
	\end{gather}
	where $\pi$ and $p_{\mu}$ are the conjugated momenta of $\varphi$ and $A_{\mu}$ respectively. The conjugated momentum $\pi_x$ is constrained to be orthogonal to $\varphi_x$, and this guarantees that $\varphi^T\varphi=1$ along the solutions of the equations of motion. Above, we omit the dependence on $\chi$ of $\mathcal{S}_{\text{simul}}$, since the pseudofermion is a spectator for the Molecular Dynamics. The symplectic integrator used to evolve the bosonic fields is a generalization of the leapfrog integrator
	\begin{gather}
		\begin{array}{l}
			\pi^{a}_{1/2}=\pi^{a}_0-\frac{\tau}{2}(\mathcal{P}^{\varphi}_{0})^{ab}\frac{\partial \mathcal{S}_{\text{simul}}}{\partial \varphi^b}(\varphi_{0},A_{0})\\
			p^{a}_{1/2,\mu}=p^{a}_{0,\mu}-\frac{\tau}{2}\frac{\partial \mathcal{S}_{\text{simul}}}{\partial A^a_{\mu}}(\varphi_{0},A_{0})\\
			\varphi^a_1=\cos(\tau|\pi_{1/2}|)\varphi^a_0+\sin(\tau|\pi_{1/2}|)\,\frac{\pi^{a}_{1/2}}{|\pi_{1/2}|}\\
			A^a_{1,\mu}=A^a_{0,\mu}+\tau p^{a}_{1/2,\mu}\\
			\pi^{a}_1=\cos(\tau|\pi_{1/2}|)\,\pi^{a}_{1/2}-\sin(\tau|\pi_{1/2}|)\,|\pi_{1/2}|\varphi^a_0-\frac{\tau}{2}\,(\mathcal{P}^{\varphi}_{1})^{ab}\,\frac{\partial \mathcal{S}_{\text{simul}}}{\partial \varphi^b}(\varphi_{1},A_{1})\\
			p^{a}_{1,\mu}=p^{a}_{1,\mu}-\frac{\tau}{2}\frac{\partial \mathcal{S}_{\text{simul}}}{\partial A^a_{1,\mu}}(\varphi_{1},A_{1}).
		\end{array}
		\label{eq:MD}
	\end{gather}
	$\mathcal{P}^\varphi_x$ is the projector on the hyperplane perpendicular to $\varphi_x$
	\begin{gather}
		(\mathcal{P}^{\varphi}_x)^{ab}=\mathbbm{1}-\varphi^a_x\,\varphi^b_x,
	\end{gather}
	The momenta $p_{x,\mu}^a$ are generated from the Gaussian distribution $P(p_{\mu})\propto e^{-p_{\mu}^2/2}$, while the momentum $\pi^a_x$ is constructed by generating an auxiliary momentum $\tilde\pi^a_x$ from the Gaussian distribution $P(\tilde\pi)\propto e^{-\tilde\pi^{2}/2}$ and by setting $\pi^a_x=\mathcal{P}_x\tilde{\pi}^a_x$. In principle, one could rewrite the equation for the momentum $\pi^a_1$ in \eqref{eq:MD} only in terms of $\phi^a_1$ using the relation between $\phi^a_0$ and $\phi^a_1$. However, we have observed that this gives rise to numerical instabilities.

	In order to compute the forces used in molecular dynamics we use Hasenbusch preconditioning \cite{HASENBUSCH2001177,PhysRevD.78.014515,Luscher:2012av}: we replace the $\mathcal K^2$ operator with $\mathcal K^2+\mu^2$ in the generation of configurations. This allows us to avoid convergence problems due to small eigenvalues fluctuating around zero. As Hasenbusch mass we use $\mu^2=\frac{1}{8\,g}$, which we have found sufficient for all values of volume, $g$, and $N$ that we considered. For the small volumes that we have considered other values of $\mu^2$ give consistent results. Larger volumes may need different values. We compute one reweighting factor accounting for the Hasenbusch preconditioning and the sign using the eigenvalues of $\mathcal{K}$
	\begin{gather}
		R=\det(\frac{1}{\sqrt{1+\mu(\mathcal{K}^{-1})^2}})\,\text{sgn}\det\mathcal{K}=\prod_i\frac{1}{\sqrt{(1+\mu^2/\lambda_i^{2})}}\,\text{sgn}\det\mathcal{K}.
	\end{gather}
	The eigenvalues are calculated using the PRIMME package \cite{PRIMME,svds_software}. In reality, we truncate this product and compute only a fraction of total eigenvalues, from the lowest up to a certain $\lambda_{i_{\text{max}}}$ for numerical stability reasons
	\begin{gather}
		R\simeq\prod_{i=1}^{i\text{max}}\frac{1}{\sqrt{(1+\mu/\lambda_i^{2})}}\,\text{sgn}\det\mathcal{K}.
	\end{gather}
	We make sure that the truncation includes all negative eigenvalues and enough eigenvalues that the truncation error is small enough. The sign of the determinant of the matrix $\mathcal{K}$ is found by counting the number of negative eigenvalues. We are currently working on improving the calculation by combining exact eigenvalues and stochastic noise vectors via deflation \cite{Luscher:2007es}. This would avoid the systematic uncertainty introduced by the truncation.

\section{Numerical results and sign problem}

	We have run simulations for the $OSp(3|2)$ and the $OSp(5|2)$-invariant models. For both theories, we have computed the bosonic and fermionic two-point function $C(t)_b\equiv\sum_{x,a}\langle\phi^a(t,x)\phi^a(0,0)\rangle$, $C(t)_f\equiv\sum_{x}\langle\psi^1(t,x)\psi^2(0,0)\rangle$ at different lattice sizes. Since the volumes that we have considered are small, we have only used point sources for the fermion disconnected contributions and we have used wall sources to do zero-momentum projection. \\
	Notice that we are interested in computing the two-point function for the original field $\phi$, which is related to the fermionic and the rescaled field $\varphi$ correlators in the following way:
	\begin{align}
		\langle\phi^a(t,x)\phi^a(0,0)\rangle&=\langle\varphi^a(t,x)\varphi^a(0,0)\rangle-\langle\varphi^a(t,x)\varphi^a(0,0)\mathcal{D}(t,x)\rangle\\&
		\nonumber-\langle\varphi^a(t,x)\varphi^a(0,0)\mathcal{D}(0,0)\rangle-\langle\varphi^a(t,x)\varphi^a(0,0)\mathcal{C}(t,x,0,0)\rangle\\&
		\nonumber +\langle\varphi^a(t,x)\varphi^a(0)\mathcal{D}(t,x)\mathcal{D}(0,0)\rangle,
	\end{align}
	where $\mathcal{D}(t,x)$ and $\mathcal{C}(t,x,0,0)$ identify the connected and disconnected components coming from the Wick contractions of the fermion fields.
	\begin{gather}
		\mathcal{D}(t,x)=\wick{\c\psi^1(t,x)\c\psi^2(t,x)}=-\mathcal{K}^{-1}_{xx}\\
		\mathcal{C}(t,x,0,0)=\wick{\c\psi^1(t,x)\c\psi^2(0,0)}\wick{\c\psi^1(0,0)\c\psi^2(t,x)}=\mathcal{K}^{-1}_{x0}\mathcal{K}^{-1}_{x0}.
	\end{gather}
	Fig.~\ref{fig:n3_l4x4} and ~\ref{fig:n5_l16x16} show the fermionic and bosonic two-point functions at three different values of $g$ for the $OSp(3|2)$ and the $OSp(5|2)$ models.\\
	In all plots, the two-point functions are compared with the corresponding correlators of the Ising or the $O(3)$-invariant model, obtained with the Swendsen-Wang and the Wolff cluster algorithm \cite{PhysRevLett.58.86,PhysRevLett.62.361}.
	
	\begin{figure}[h]
		\centering
		\includegraphics[width=1\linewidth]{n3_l4x4.eps}
		\caption{Two-point functions $C(t)$ for the bosonic (in red) and fermionic (in blue) fields of the $OSp(3|2)$ model on a $4\times4$ lattice, expressed in lattice spacing units. The black dotted line represents the correspondent two-point correlator for the Ising model.}
		\label{fig:n3_l4x4}
	\end{figure}
	\begin{figure}[h]
		\includegraphics[width=\textwidth]{n5_l16x16.eps}
		\caption{Two-point functions $C(t)$ for the bosonic (in red) and fermionic (in black) fields of the $OSp(5|2)$ model on a $16\times16$ lattice, expressed in lattice spacing units. Here the black dotted line represents the correspondent two-point correlator for the $O(3)$ model.}
		\label{fig:n5_l16x16}
	\end{figure}
\begin{figure}
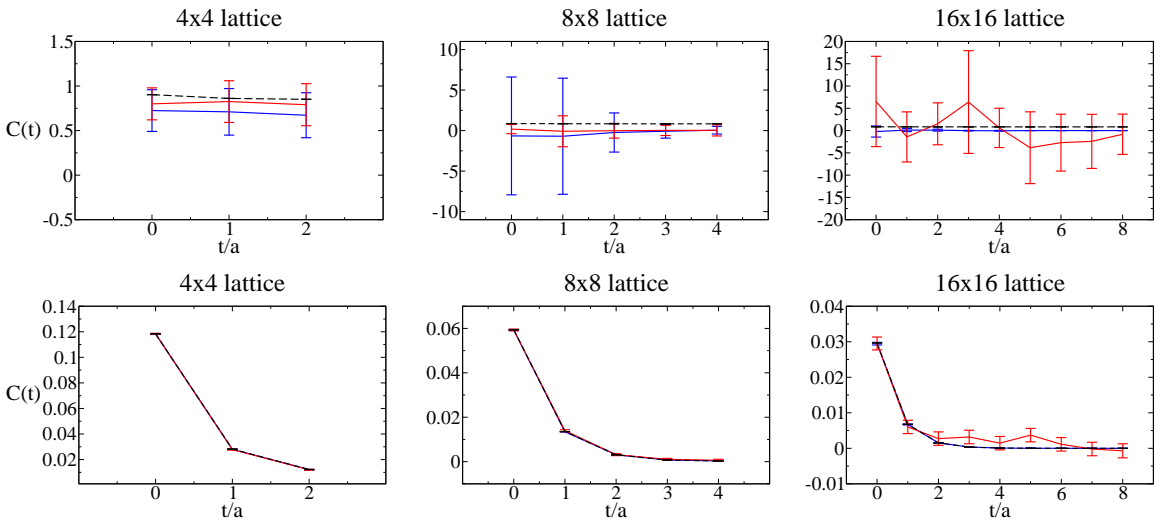

		\includegraphics[width=1\linewidth]{n3g4_vol.eps}
		
		\vspace{0.2cm}
		
		\includegraphics[width=1\linewidth]{n5g4_vol.eps}
		\caption{Plot of the correlators $C(t)$ of the $OSp(3|2)$ and $OSp(5|2)$ model at same value of the coupling $g=4.0$. From the plots, it's easy to notice that the effects of the sign problem on the correlators are less visible in the $OSp(5|2)$ models even on bigger lattice sizes.}
		\label{fig:sign_with_vol}
	\end{figure}

	We observe that the correlators are equal within $2\sigma$ and thus behave as predicted from the analytic results in equation \eqref{eq:equi:n-pt}. However, for smaller values of the coupling and higher lattice sizes, the large statistical errors due to the sign problem render the results not a significant. The sign problem appears to be more severe in the $OSp(3|2)$ model as is evident in Fig. \ref{fig:n3_l4x4} and Fig. \ref{fig:sign_with_vol}, where we expect that $\langle s \rangle\rightarrow1$ only for very large values of $g$. On the other hand in the $OSp(5|2)$ case for $g\gtrsim7$ and for all the lattice volumes that we considered, the expectation value of the sign is approximately 1. The fact that the sign problem improves in theories with a higher number of bosons can be predicted from the form of the $\mathcal{K}$ matrix in \eqref{eq:K_def}: as the term proportional to $N$ grows, the spectrum of the matrix becomes more and more positive. To further illustrate the effects of the sign problem on the correlators, we plot the correlators of the $OSp(3|2)$ and $OSp(5|2)$ models at a certain value of the coupling in Figure \ref{fig:sign_with_vol}. From the plots, it is evident that the effects of the sign problem on the correlators are less visible in the $OSp(5|2)$ models, even on larger lattice sizes.

\begin{figure}[H]
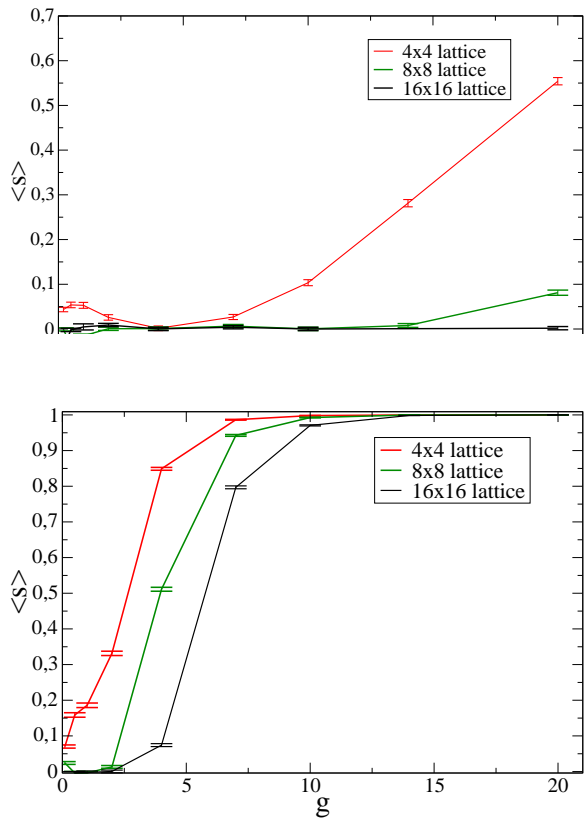

		\centering
		\includegraphics[width=0.5\textwidth]{sign_n3.eps}

		\vspace{0.2cm}

		\includegraphics[width=0.5\textwidth]{sign_n5.eps}
		\caption{Values of $\langle s\rangle$ computed for the $OSp(3|2)$ (top figure) and the $OSp(5|2)$ (bottom figure) as a function of the coupling $g$.}
		\label{sign_plot}
	\end{figure}

	The behavior of $\langle s \rangle$ as a function of the coupling and the lattice volume is shown in Fig. \ref{sign_plot}. Even though the number of points is limited, we have tried some simple fits of the different $\langle s \rangle$ with the volume and the coupling. The fit appears consistent with an exponential behavior, decreasing with the volume $V$ or the inverse of the coupling \cite{PhysRevLett.83.3116}. Due to the severe sign problem in the current HMC setting, exploring the range of physical interest, for example, the phase transition at $g \sim 4$ of the $OSP(3|2)$ model, appears impractical. \\

	\section{Outlook}
	
	In this work, we have constructed a discretized action and an algorithm for simulating the $OSp(N+2|2)$ non-linear sigma model. We have presented numerical results for the bosonic and fermionic two-point functions of the $OSp(3|2)$ and $OSp(5|2)$ sigma models.

	We have observed the emergence of a sign problem in the simulations. This sign problem comes from $\mathcal K$ admitting negative eigenvalues. For smaller couplings, some eigenvalues appear to fluctuate more and more around zero. The sign problem seems to be milder in the $OSp(5|2)$ case, with the mean value of the sign approaching $1$ for higher values of the coupling.
	We intend to compute other observables, like the four-point functions and the conserved currents, and to study the behavior of the sign problem in more detail~\cite{inprogress}. 

	In future work, we intend to explore algorithms that treat the fermions differently, hoping to ameliorate the sign problem.

	\section{Acknowledgements}
	We thank Mika Lauk for sharing his code for the Wolff cluster algorithm.  
	The research of I. C. and J. H. W. is funded by the Deutsche Forschungsgemeinschaft (DFG,
	German Research Foundation) - Projektnummer 417533893/GRK2575 ”Rethinking Quantum Field Theory”. The research of V.F. is supported by the Heisenberg Professorship program  506208580 ``Quantum strings and gauge fields at arbitrary coupling''.
	
	\bibliographystyle{JHEP}
	\bibliography{europlex}
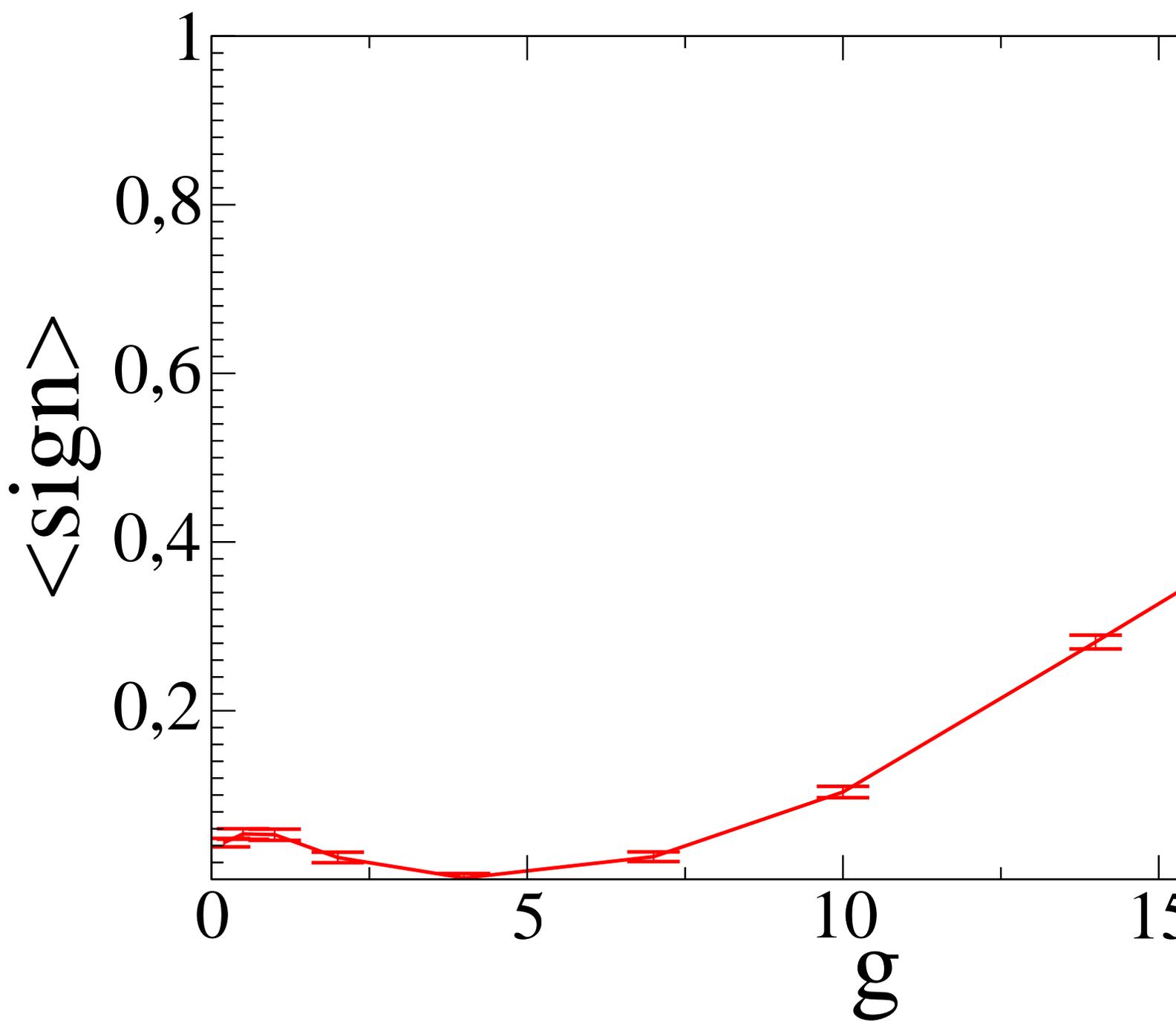
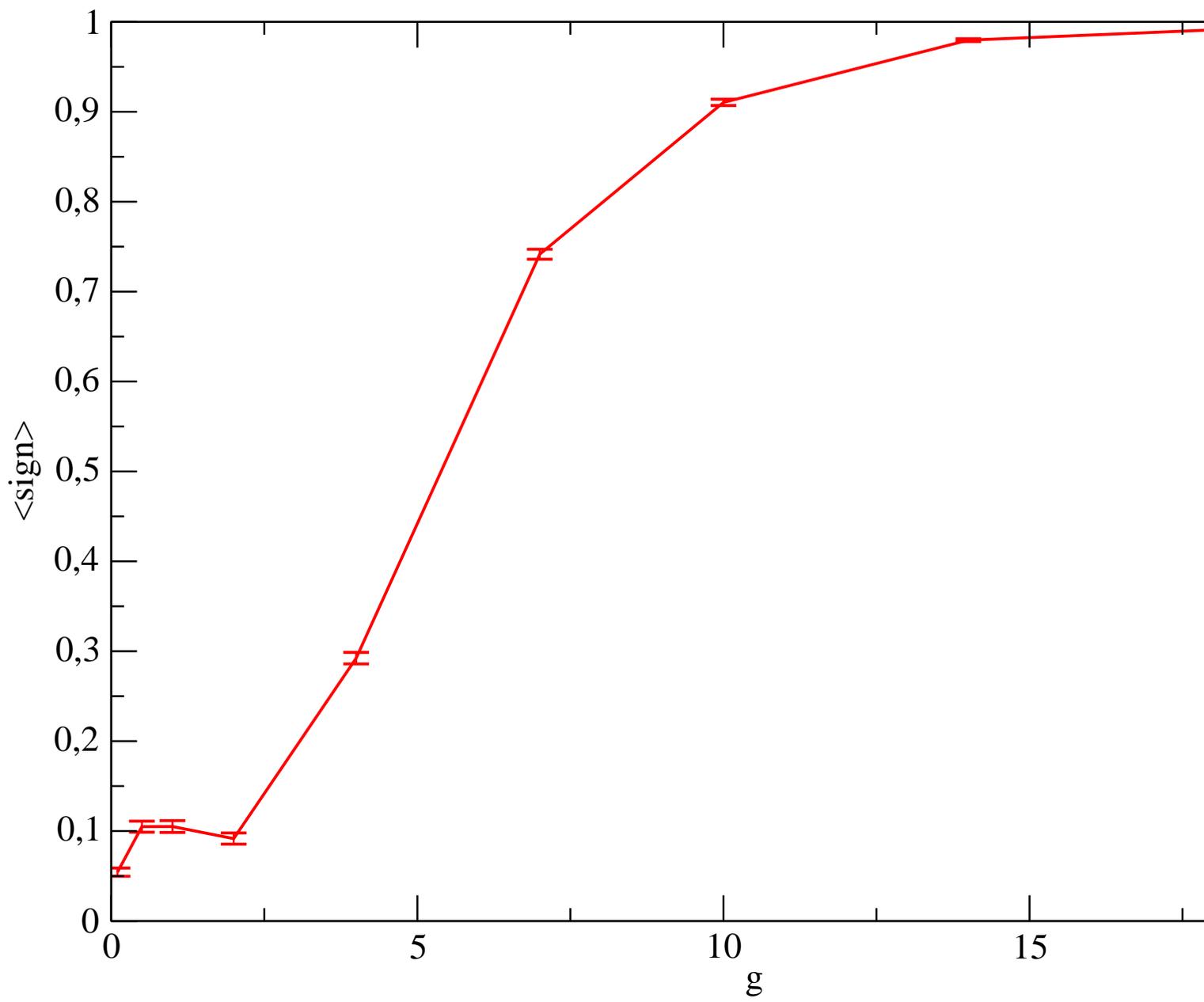
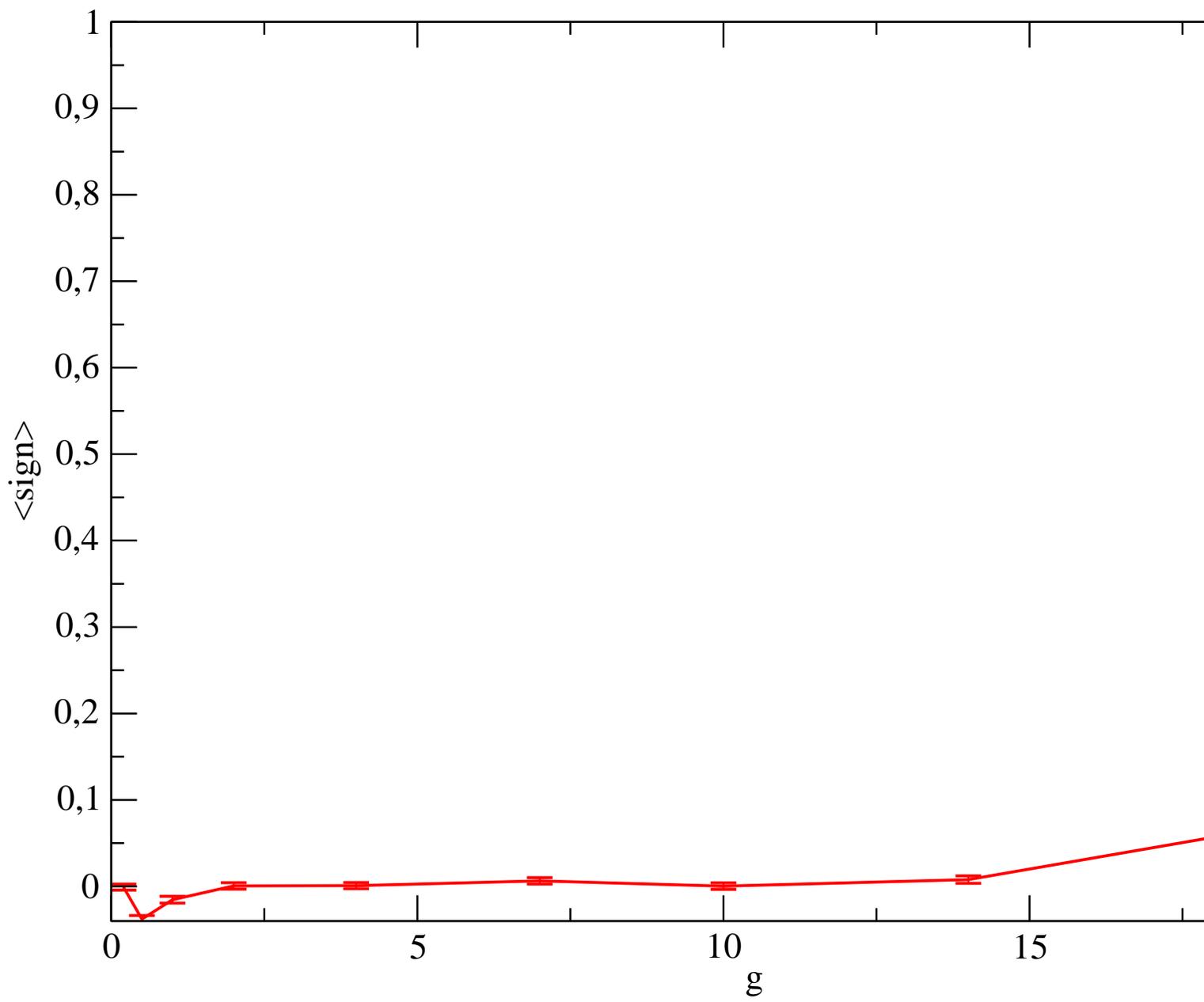
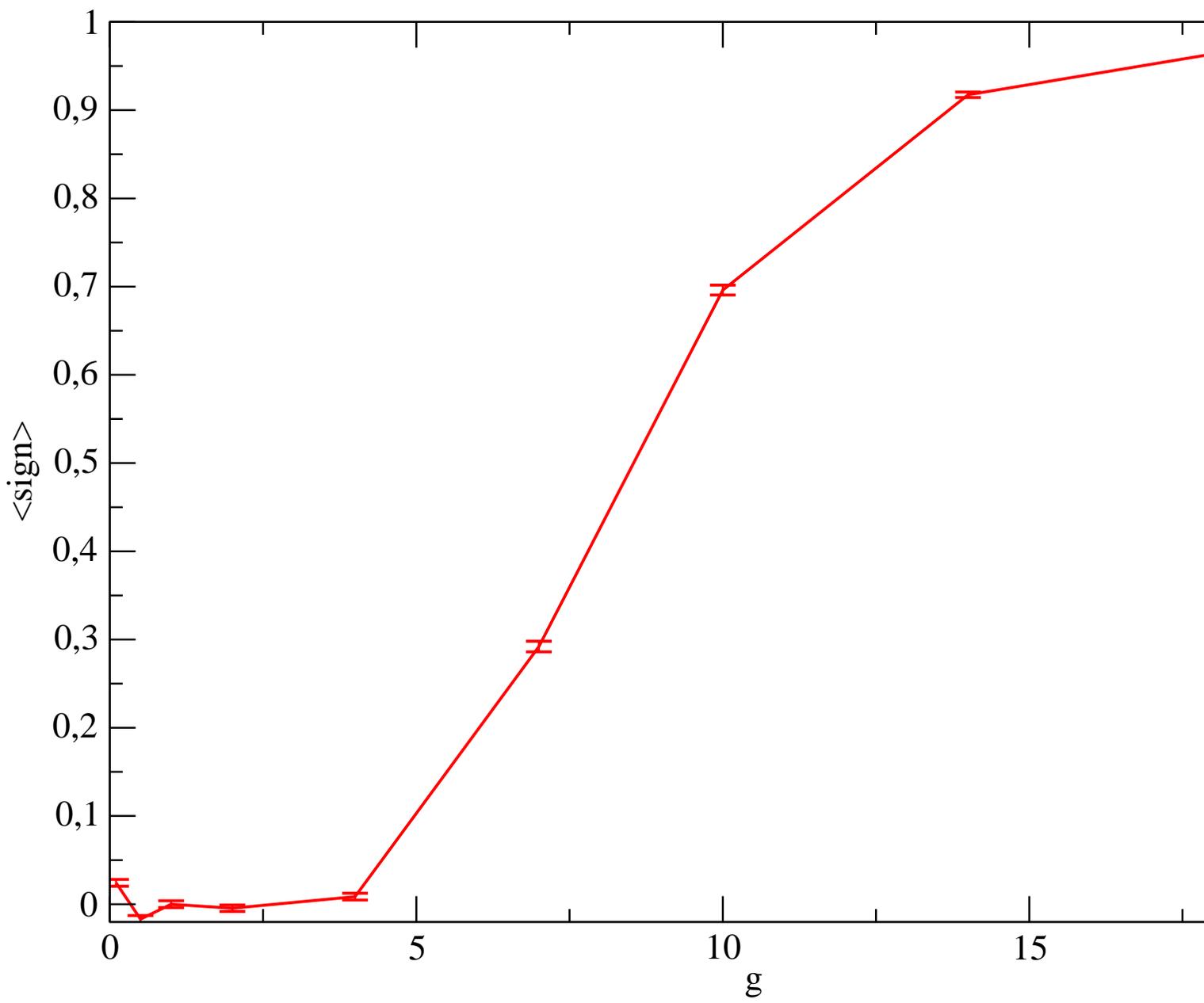
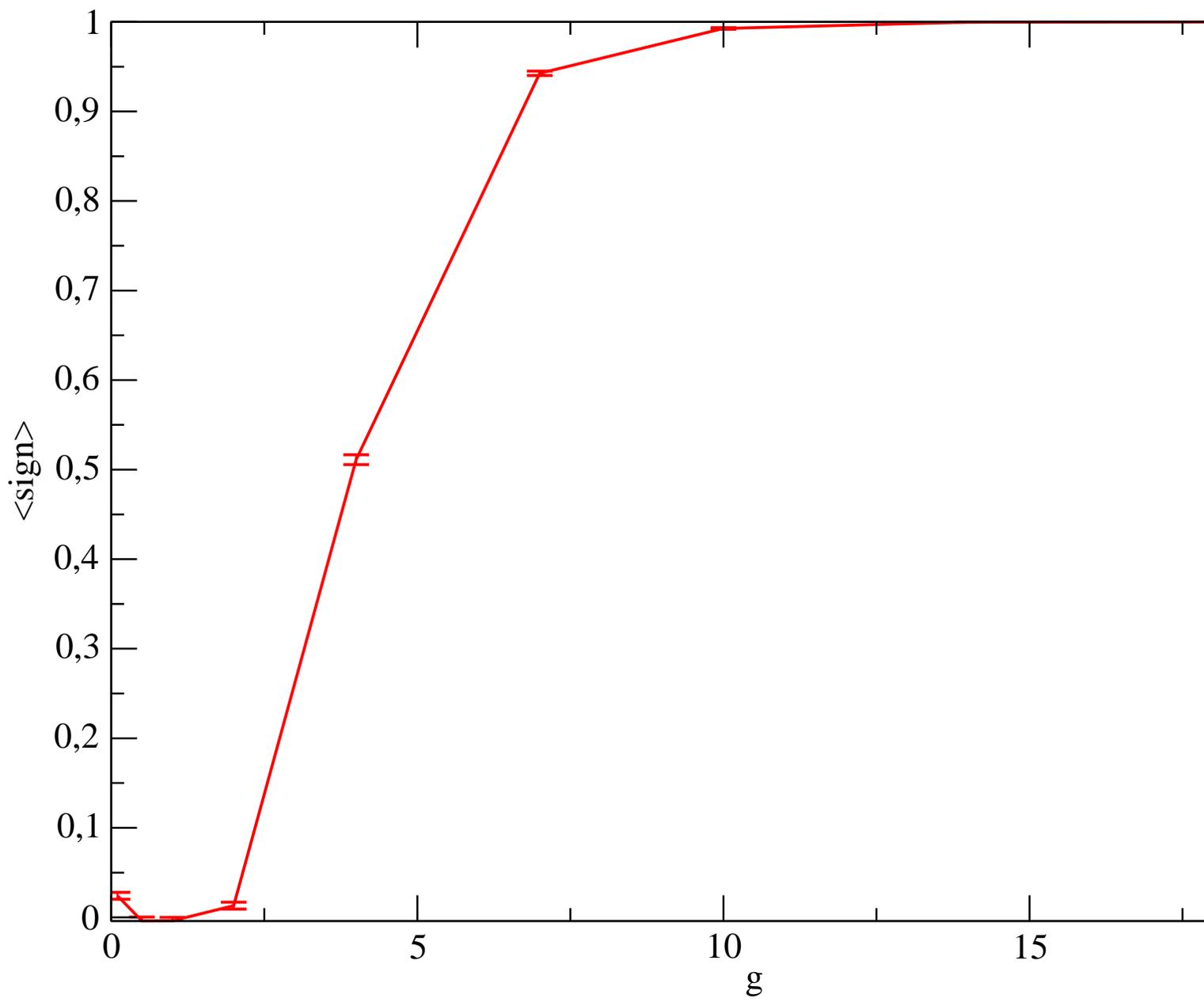
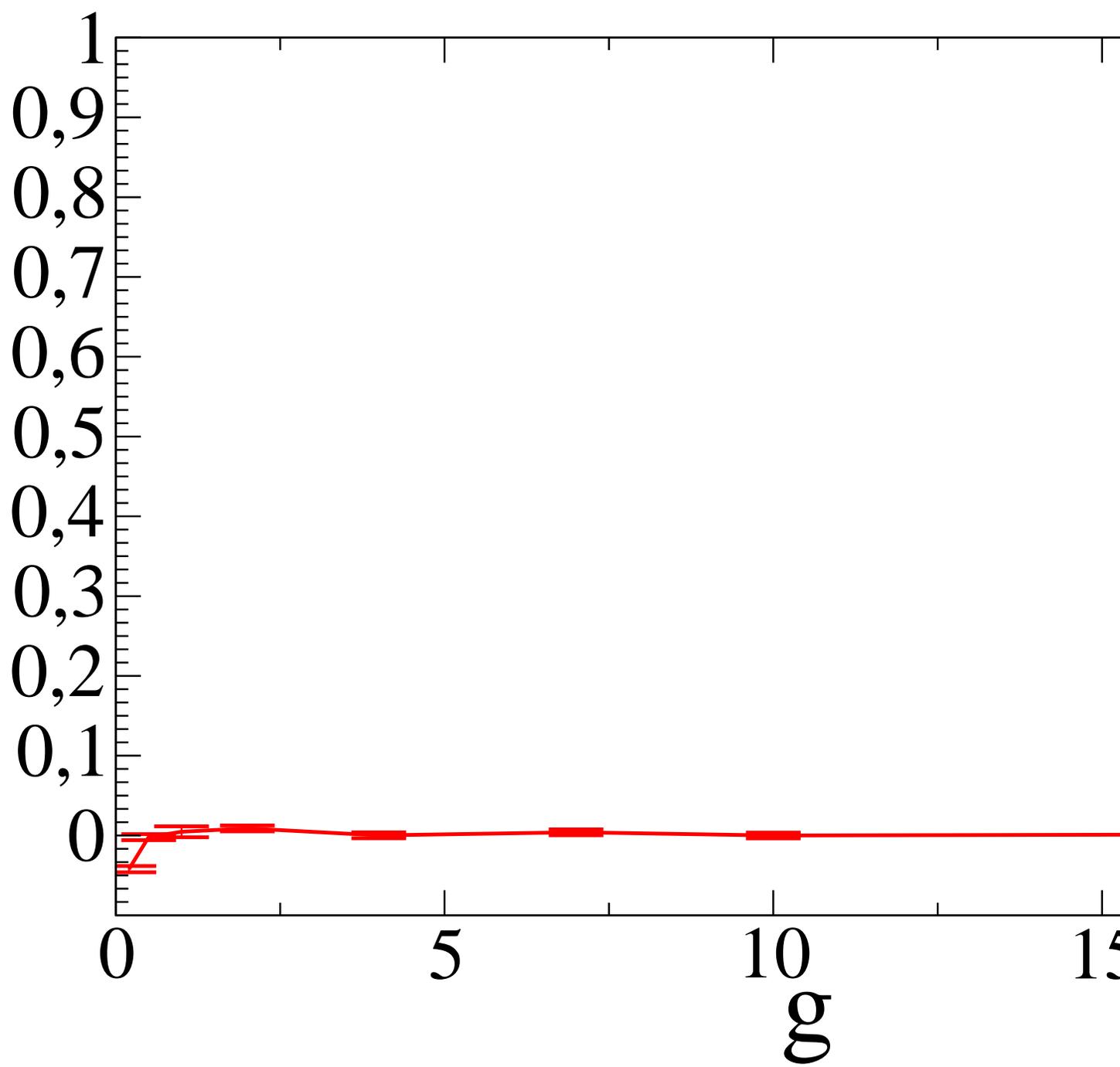
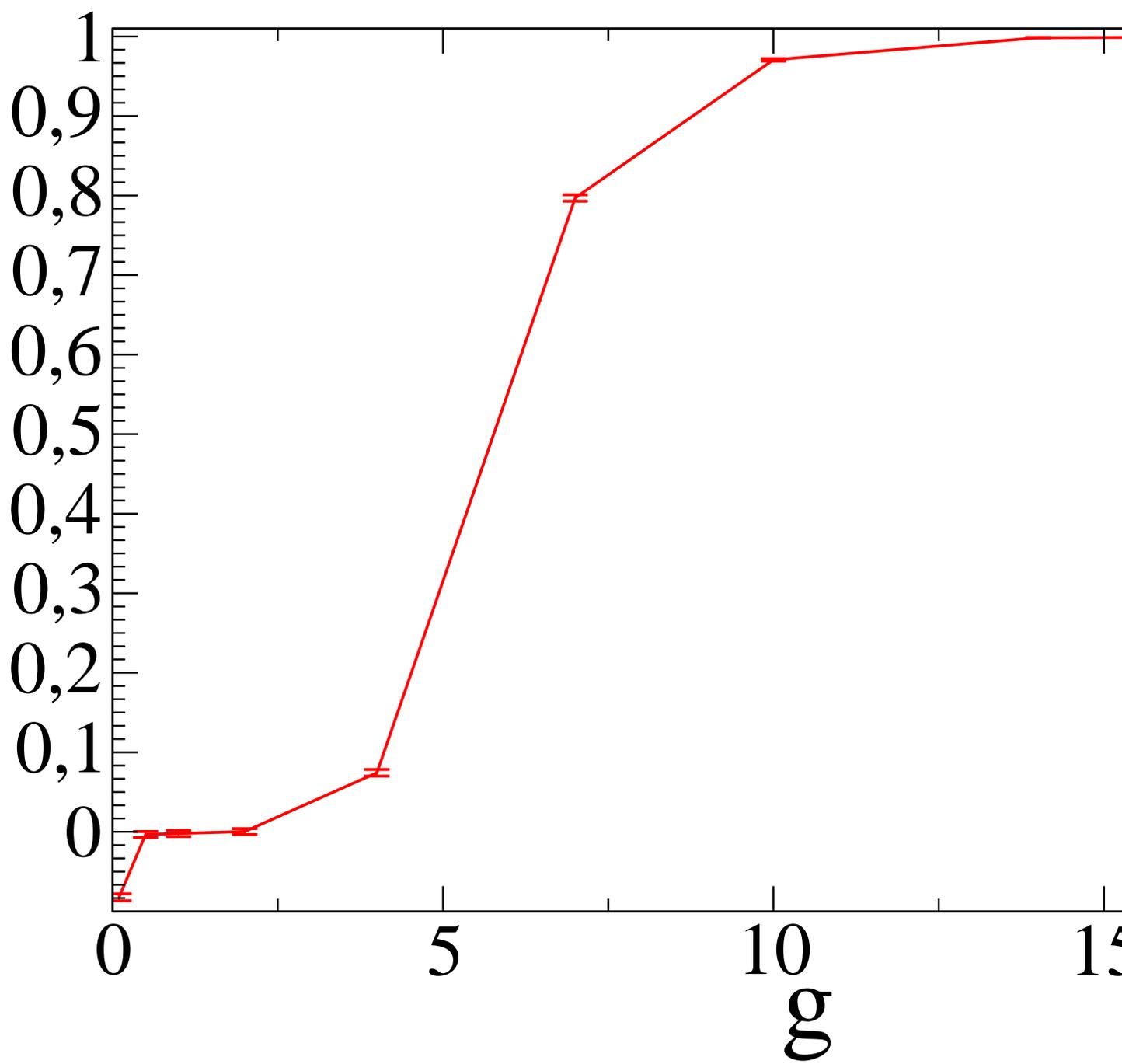
\end{document}